\documentclass{kluwer}
\usepackage{graphicx}
\newcommand{\EQ}{\begin{equation}}
\newcommand{\EN}{\end{equation}}
\newcommand{\eq}[1]{(\ref{#1})}
\newcommand{\EEq}[1]{Equation~(\ref{#1})}
\newcommand{\Eq}[1]{Eq.~(\ref{#1})}
\newcommand{\Eqs}[2]{Eqs.~(\ref{#1}) and~(\ref{#2})}
\newcommand{\eqs}[2]{(\ref{#1}) and~(\ref{#2})}
\newcommand{\Eqss}[2]{Eqs.~(\ref{#1})--(\ref{#2})}
\newcommand{\eqss}[2]{(\ref{#1})--(\ref{#2})}
\newcommand{\Sec}[1]{Section~\ref{#1}}
\newcommand{\Fig}[1]{Figure~\ref{#1}}

\newcommand{\Tab}[1]{Table~\ref{#1}}

\newcommand{\Figs}[2]{Figs~\ref{#1} and \ref{#2}}
\newcommand{\ea}{{\em et al., }}
\def\half{{\textstyle{1\over2}}}
\def\quarter{{\textstyle{1\over4}}}
\newcommand{\dd}{{\rm d} {}}
\newcommand{\ysci}[5]{: #1, #5, {\em Science }{\bf #2}, #3--#4.}
\newcommand{\ynat}[5]{: #1, #5, {\em Nature }{\bf #2}, #3--#4.}
\newcommand{\yprl}[5]{: #1, #5, {\em Phys.\ Rev.\ Lett. }{\bf #2}, #3--#4.}
\newcommand{\yphy}[5]{: #1, #5, {\em Physica }{\bf #2}, #3--#4.}
\newcommand{\yphl}[5]{: #1, #5, {\em Phys.\ Lett. }{\bf #2}, #3--#4.}
\newcommand{\yoleb}[5]{: #1, #5, {\em Orig.\ Life Evol.\ Biosph. }{\bf #2}, #3--#4.}
\newcommand{\yjour}[6]{: #1, #6, {\em #2} {\bf #3}, #4--#5.}
\newcommand{\ybook}[3]{: #1, {\em #2}, #3.}

\begin{document}
\begin{article}
\begin{opening}
\title{Homochiral growth through enantiomeric cross-inhibition}

\author{A. \surname{Brandenburg}$^*$}
\author{A.~C. \surname{Andersen}}
\author{S. \surname{H\"ofner}}
\author{M. \surname{Nilsson}}
\institute{Nordita, Blegdamsvej 17, DK-2100 Copenhagen \O, Denmark\\
($*$ author for correspondence, e-mail: brandenb@nordita.dk,
phone +45 353 25228, fax: +45 353 89157)}

\runningtitle{Homochiral growth through enantiomeric cross-inhibition}
\runningauthor{Brandenburg et al.}

\begin{abstract} 
The stability and conservation properties of a recently
proposed polymerization model are studied.
The achiral (racemic) solution is linearly unstable once the
relevant control parameter (here the fidelity of the catalyst)
exceeds a critical value.
The growth rate is calculated for different fidelity parameters and
cross-inhibition rates.
A chirality parameter is defined and shown to
be conserved by the nonlinear terms of the model.
Finally, a truncated version of the model
is used to derive a set of two ordinary
differential equations and it is argued that these equations are more
realistic than those used in earlier models of that form.
\end{abstract}


\end{opening}

\section{Introduction}

The chirality of molecules in living organisms must have been fixed at
an early stage in the development of life.
All life that we know is based on RNA and DNA molecules with
dextrarotatory sugars.
There is growing evidence that the RNA world
(Woese, 1967; Crick, 1968; Orgel, 1968; see also Wattis \& Coveney 1999)
must have been preceded by a simpler pre-RNA world made up of
achiral constituents (Bada, 1995, Nelson \ea 2000).
An alternative carrier of genetic code are peptide nucleic acids or PNA
(Nielsen, 1993).
These can be rather simple and are currently discussed in connection
with the idea to build artificial life (Rasmussen \ea 2003).
Furthermore, although PNA can still be chiral (Tedeschi \ea 2002),
there are also forms of PNA that are achiral (Pooga \ea 2001),
suggesting that chirality may have
developed later when the first RNA molecules formed.

In current proposals to build artificial life, chirality does not
seem to be crucial.
The PNA molecules is proposed to act primarily as charge carrier, 
i.e.\ a very primitive functionality compared to the genetic code
in contemporary cells (Rasmussen \ea 2003). 
At this stage, homochirality may have been introduced by chance.
This is also supported by the fact that
chiral polymers of the same chirality tend to have a more stable
structure (Pogodina \ea 2001) and would therefore be
genetically preferred.

Since the introduction of chiral molecules is assumed to take place
at a stage when there is already growth and self-replication, it is
also plausible to assume that the existence of chiral molecules has
an autocatalytic effect in producing new chiral molecules of the
same chirality (Kondepudi \ea 1990).
This is the basis of the recently proposed polymerization model of
Sandars (2003); see also Wattis \& Coveney (2004).
The purpose of the present paper is to reconsider this model
(or a slightly modified version of it) and to analyze its
stability behavior and conservation properties.
We also discuss and illustrate some of the
salient features of the model in more detail.
The model is then compared with earlier models of homochirality where
the detailed polymerization process is ignored and the dynamics of single
variables representing left and right handed polymers are modeled instead
(Frank, 1953; Kondepudi and Nelson, 1984; Goldanskii and Kuzmin 1989;
Avetisov and Goldanskii, 1993; Saito and Hyuga, 2004).

In order to appreciate the nature of the many terms in the model of Sandars we
begin by discussing first the basic principle of the model in connection
with homochiral polymer growth and then turn to the full set of reactions
that are included in the model.

\section{Homochiral polymer growth}
\label{HomochiralGrowth}

In this section we discuss the growth of polymers by adding monomers
of the same chirality, i.e.\ we ignore reactions with monomers of the
opposite chirality.
This is conceptually the simplest case, but its equilibrium solution
also corresponds to a solution of the full system discussed below.
We write down the equations for left-handed polymers, but the same
applies also to right-handed polymers.

A left-handed polymer of length $n$ is assumed to react with
a left-handed monomer via the reaction
\begin{equation}
L_{n}+L_1\rightarrow L_{n+1}\quad\mbox{($n\ge1$)}.
\end{equation}
The reaction rate is $k_S$, but since $L_{n}$ can bind to $L_1$
on either side, the total reaction rate is $2k_S$ and proportional
to the product of the concentrations of the two constituents.
We denote the concentration of $L_n$ chains by $[L_n]$, so
in a volume $V$ the number of $L_n$ chains is $N_n\equiv[L_n]V$.
For $n\geq3$ the number of possible pairs of $L_{n-1}$ and $L_1$ is
$N_n\times N_1$.
A special situation arises for $n=2$, because then $L_1$ is interacting
with another $L_1$, and the number of possible pairs is only
$\half N_1(N_1-1)\approx\half N_1^2$.
[This problem is familiar from the physics of nuclear reactions; see, e.g.,
Kippenhahn and Weigert (1990).]
We therefore introduce the factor $\sigma_n^{(1/2)}$ defined by
\begin{equation}
\sigma_n^{(\alpha)}=\left\{
\begin{array}{cc}
\alpha & \quad\mbox{for}\quad n=2,\\
1 & \quad\mbox{for}\quad n\ge3.
\end{array}
\right.
\label{sigman}
\end{equation}
(Later we shall use this factor also with $\alpha=0$ instead of $1/2$.)
The corresponding contribution to the evolution of the concentration of
$L_n$ is therefore
\begin{equation}
{\dd[L_{n+1}]\over\dd t}=...+2k_S\sigma_n^{(1/2)}[L_{n}][L_1],
\end{equation}
where the dots denote the presence of other terms that will be discussed
later.

Obviously, the concentrations of $L_{n}$ and $L_1$ have to decrease
at the same rate by the same amount, so
\begin{equation}
{\dd[L_{n}]\over\dd t}=...-2k_S\sigma_n^{(1/2)}[L_{n}][L_1],
\label{Lna}
\end{equation}
\begin{equation}
{\dd[L_1]\over\dd t}=...-2k_S\sigma_n^{(1/2)}[L_{n-1}][L_1].
\label{L1a}
\end{equation}
In the following we regard 
$n$ as a general index with $2\leq n\leq N$, the evolution
of $[L_n]$ is governed by the difference of two terms (gain from
$L_{n-1}$ chains and loss in favor of producing $L_{n+1}$ chains).
The production of each $L_n$ contributes to a loss of $L_1$ monomers,
so the right hand side of \Eq{L1a} becomes a sum over all $n$.
The full set of equations is then
\begin{equation}
{\dd[L_1]\over\dd t}=Q_L-\lambda_L[L_1],
\quad\mbox{where}\quad\lambda_L=2k_S\sum_{n=1}^{N-1}[L_{n}],
\label{L1b}
\end{equation}
\begin{equation}
{\dd[L_{n}]\over\dd t}=2k_S[L_1]\Big(\sigma_n^{(1/2)}[L_{n-1}]-[L_{n}]\Big),
\label{Lnb}
\end{equation}
where $Q_L$ denotes the production of new $L_1$ monomers (see below).
A corresponding set of equations applies also to right-handed polymers,
i.e.\ $R_1$ and $R_n$.
Note that \Eqs{L1b}{Lnb} obey the conservation law
\begin{equation} 
{\dd E_L\over\dd t}=Q_L-2k_S[L_1][L_N],
\label{dELdt}
\end{equation}
where
\begin{equation}
E_L=\sum_{n=1}^N n[L_n]
\label{ELdef}
\end{equation}
is the total number of left-handed building blocks.
This number reaches an equilibrium if the supply of
new left-handed monomers, $Q_L$, balances the loss associated
with reactions involving the longest polymers possible for a
given value of $N$.

\EEq{Lnb} shows that in the steady state we have $[L_n]=\half[L_1]$
for all $n\geq2$.
Using \Eq{L1b}, we find $\lambda_L=k_S N[L_1]$, and therefore
\begin{equation}
2[L_n]=[L_1]=\sqrt{Q_L/k_SN}\quad\mbox{(steady state)}
\label{HomochiralEquilib}
\end{equation}
is a possible equilibrium solution.

New left and right handed
monomers are assumed to be continuously reproduced from an
achiral (racemic) substrate.
The rates of regeneration, $Q_L$ and $Q_R$, depend on the concentration of the
substrate, $[S]$, and in some fashion on the relative concentrations
of right and left handed polymers.
So, in general, we write
\begin{equation}
Q_L=k_C[S]\Big\{\half(1+f)C_L+\half(1-f)C_R+C_{0L}\Big\},
\label{QLdef}
\end{equation}
\begin{equation}
Q_R=k_C[S]\Big\{\half(1+f)C_R+\half(1-f)C_L+C_{0R}\Big\},
\label{QRdef}
\end{equation}
where $C_L$ and $C_R$ are some measures of the catalytic effect of the
already existing right and left handed polymers, and the terms $C_{0L}$
and $C_{0R}$ allow for the possibility of non-catalytic production of
left and right handed monomers -- possibly at different rates.
(Unless noted otherwise, we keep $C_{0L}=C_{0R}=0$.)

The concentration of the substrate is assumed to be maintained by
a source $Q$, so we have
\begin{equation}
{\dd[S]\over\dd t}=Q-\left(Q_L+Q_R\right),
\label{dSdt}
\end{equation}
where $Q_L+Q_R=k_C[S](C_L+C_R+C_{0L}+C_{0R})$; see \Eqs{QLdef}{QRdef}.
In general, we expect $C_L$ and $C_R$ to be some function of
$L_n$ and $R_n$, respectively.
Sandars (2003) assumed $C_L=[L_N]$ and $C_R=[R_N]$, i.e.\ the
catalytic effect depends on the concentrations of the longest possible
chains of left and right handed polymers.
This assumption imposes a dependence on the cutoff value $N$,
a dependence that should preferably be avoided in numerical or other 
technical considerations.
The model should for example be stable and consistent in the
limit when $N$ is infinite.
Another option would be to assume $C_L=[L_M]$ and $C_R=[R_M]$, where $M<N$
is a fixed value that is independent of the maximum chain length.
Both alternatives have the disadvantage that $[L_1]$ and $[R_1]$ can never
grow unless $[L_M]$ or $[R_M]$ are initially also finite.
While it is plausible that long chains carry more catalytic weight than shorter
ones, the dependence of the results on the particular choice of $M$
seems artificial.
(The allowance of finite values of $C_{0L}$ and $C_{0R}$ would remove
this problem, although in practice both of these values should still be quite small.)

On may expect that the catalytic properties of the existing left
and right handed polymers depend on the length of the
polymer. The exact functional expression for this dependence
is not known. It is therefore important that a model that explains
homochirality is not sensitive to the details of the catalytic
properties and hence the functional form of $C _L$ and $C _R$.
It turns out that the qualitative
behavior of the model of Sandars is indeed robust in this respect,
e.g.\ a pitchfork bifurcation exists in both Sandars' original
and in our model.
To avoid artificial dependence on the maximal chain length
$N$, we chose to let the catalytic functions have the following
form 
\begin{equation}
C_L=E_L,\quad
C_R=E_R,
\label{Cdef}
\end{equation}
where $E_L$ is given by \Eq{ELdef}, and $E_R$ is defined analogously.
This is similar to the choice of Wattis \& Coveney (2004) who assumed,
independently of us, $C_L=E_L-[L_1]$ and $C_R=E_R-[R_1]$.

We now comment on another aspect of the model of Sandars.
He assumed that in the evolution of $[L_N]$ the loss is not
$2k_S[L_1][L_N]$, as it would be if \Eq{Lnb} were applied to $n=N$,
but he introduced an explicit linear damping term instead.
This implies that the model behaves discontinuously at the end
of the chain.
We feel that an ``extrapolating'' (continuous) behavior
is more reasonable, so we choose to apply \Eq{Lnb} also at $n=N$.

\begin{figure}[t!]\begin{center}
\includegraphics[width=\textwidth]{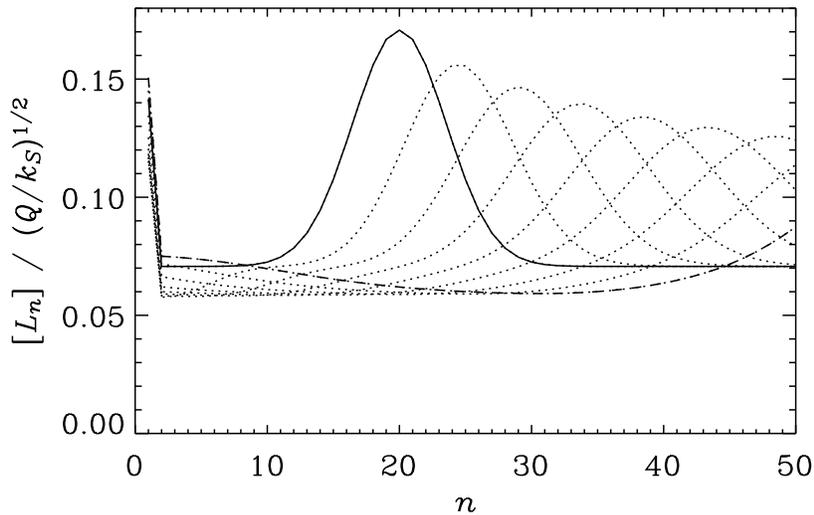}
\end{center}\caption[]{
Wave-like propagation of a finite amplitude perturbation.
The initial profile is a gaussian.
Note the undisturbed propagation of the wave out of the chain at $n=N$.
The time difference between the different curves is $20/(k_SQ)^{1/2}$.
We have shown the first and last times as solid and dashed lines,
and all other times as dotted lines.
The parameters are $N=50$ and $k_C/k_S=1$.
}\label{pwave}\end{figure}

It is interesting to note that in the continuous limit, \Eq{Lnb}
becomes
\begin{equation}
\left({\partial\over\partial t}+2k_S[L_1]
{\partial\over\partial n}\right)[L_n]=0,
\end{equation}
which describes waves
traveling toward larger $n$.
This is shown in \Fig{pwave}, where we have perturbed
the equilibrium solution \eq{HomochiralEquilib} by a gaussian
and have solved \Eqss{L1b}{Lnb} numerically.
The wave is damped and has a speed that is proportional to $(k_SQ)^{1/2}$,
because for the steady state background solution $[L_1]\sim (k_S/Q)^{1/2}$.
Note that the extrapolating boundary condition at $n=N$ allows
the wave to escape freely.

In this paper we do not adopt the nondimensionalization of Sandars.
Instead, we note that there are only two physical dimensions in this
problem: time and volume.
Characteristic quantities with the dimensions of time and volume are
$(k_SQ)^{-1/2}$ and $(k_S/Q)^{1/2}$, respectively.
We therefore present all results by explicitly quoting these dimensions.
In practice this means that from now on we use $Q=k_S=1$ as numerical
values, but we keep the symbols in the equations for clarity.
Throughout this paper we also assume $k_C/k_S=1$; calculations with
different values do not seem to affect our results in any important way.

The fact that the equilibrium solution is constant for all $n\geq2$
implies that this value will decrease for longer choices of $N$.
In that sense the solution is never converged.
This situation changes when we allow the ends of the left-handed polymers
to be spoiled by right-handed monomers, as done by Sandars (2003).
This will be discussed in the next section.

\section{Enantiomeric cross-inhibition}

Already 20 years ago, Joyce \ea (1984) showed in an important paper
describing experiments with template-directed polymerization that,
once a monomer of the opposite chirality is bound to one end of the
chain, the polymerization terminates on that end of the chain.
Sandars (2003) incorporated this effect in his model and showed that
this can lead to a bifurcation into two possible solutions of opposite
chirality and hence to homochirality.

The full set of reactions included in his model is (for $n\ge2$)
\begin{eqnarray}
L_{n}+L_1&\stackrel{2k_S~}{\longrightarrow}&L_{n+1},
\label{react1}\\
L_{n}+R_1&\stackrel{2k_I~}{\longrightarrow}&L_nR_1,
\label{react2}\\
L_1+L_{n}R_1&\stackrel{k_S~}{\longrightarrow}&L_{n+1}R_1,
\label{react3}\\
R_1+L_{n}R_1&\stackrel{k_I~}{\longrightarrow}&R_1L_nR_1,
\label{react4}
\end{eqnarray}
and for all four equations we have the complementary reactions
obtained by exchanging $L\rightleftarrows R$.
Following Sandars (2003), we have introduced the new parameter $k_I$,
which quantifies that rate of enantiomeric cross-inhibition.
The special case $k_I=0$ corresponds to the case discussed in the
previous section.

The most important effect of enantiomeric cross-inhibition is that
a certain fraction of chains becomes spoiled by producing $L_nR_1$
and $R_nL_1$ polymers.
\EEq{Lnb} and its complementary equation for right-handed polymers
suffer therefore a loss proportional to $2k_I$, so we have instead
\begin{equation}
{\dd[L_{n}]\over\dd t}=2k_S[L_1]\Big(\sigma_n^{(1/2)}[L_{n-1}]-[L_{n}]\Big)
-2k_I[L_{n}][R_1],
\label{Ln_new2}
\end{equation}
\begin{equation}
{\dd[R_{n}]\over\dd t}=2k_S[R_1]\Big(\sigma_n^{(1/2)}[R_{n-1}]-[R_{n}]\Big)
-2k_I[R_{n}][L_1].
\label{Rn_new2}
\end{equation}
These equations allow us to see what happens in the racemic case
with $[R_n]=[L_n]$.
In a steady state we have (for $n\ge2$)
\begin{equation}
[L_n]=\half a^{-(n-1)}[L_1]
\quad\mbox{(racemic solution)},
\end{equation}
where we have defined $a=1+k_I/k_S$.
In particular, if $k_I=k_S$, then $[L_n]=2^{-n}[L_1]$,
i.e.\ $[L_n]$ drops by a factor of 2 from one $n$ to the next,
except for $n=1$ to 2, where it drops by a factor of 4.
We should note, however, that this solution can be unstable
(see \Sec{SStability}).

So far, we have not yet considered the evolution equations for the
concentrations of the mixed terms, $L_{n}R_1$ and $R_{n}L_1$.
Following Sandars (2003), we abbreviate the corresponding
concentrations by $[L_{n}R]$ and $[R_{n}L]$, respectively, i.e.\ without
the subscript 1 on the terminating end of the chain.
The effect of generating these terms was already manifested in
\Eqs{Ln_new2}{Rn_new2} through the appearance of the last term proportional
to $2k_I$.
Nevertheless, we do need to solve for $[L_{n}R]$ and $[R_{n}L]$
explicitly, because the reactions \eqs{react3}{react4} consume
$L_1$ and $R_1$ monomers, respectively.
The evolution equations for $[L_1]$ and $[R_1]$ are therefore given by
\begin{equation}
{\dd[L_1]\over\dd t}=Q_L-\lambda_L[L_1],\quad
{\dd[R_1]\over\dd t}=Q_R-\lambda_R[R_1],
\label{L1R1}
\end{equation}
where
\begin{equation}
\lambda_L=
2k_S\sum_{n=1}^{N-1}[L_n]
+2k_I\sum_{n=1}^{N}[R_n]
+k_S\sum_{n=2}^{N-1}[L_nR]
+k_I\sum_{n=2}^{N}[R_nL],
\label{lambdaL}
\end{equation}
\begin{equation}
\lambda_R=
2k_S\sum_{n=1}^{N-1}[R_n]
+2k_I\sum_{n=1}^{N}[L_n]
+k_S\sum_{n=2}^{N-1}[R_nL]
+k_I\sum_{n=2}^{N}[L_nR],
\label{lambdaR}
\end{equation}
are the decay rates that quantify
the losses associated with the reactions
\eqss{react1}{react4}, respectively.

In \Eqs{lambdaL}{lambdaR} the concentrations $[L_{n}R]$ and $[R_{n}L]$
enter, so we have to solve their corresponding
evolution equations (for $n\geq2$)
\begin{equation}
{\dd[L_{n}R]\over\dd t}=
k_S[L_1]\Big(\sigma_n^{(0)}[L_{n-1}R]-[L_{n}R]\Big)
+k_I[R_1]\Big(2[L_{n}]-[L_nR]\Big),
\label{LRevol}
\end{equation}
\begin{equation}
{\dd[R_{n}L]\over\dd t}=
k_S[R_1]\Big(\sigma_n^{(0)}[R_{n-1}L]-[R_{n}L]\Big)
+k_I[L_1]\Big(2[R_{n}]-[R_nL]\Big),
\label{RLevol}
\end{equation}
where the $\sigma_n^{(0)}$ factor turns off the first term for $n=2$;
see \Eq{sigman}.
In \Eqs{LRevol}{RLevol}
the first two terms proportional to $k_S$ correspond to
the homochiral growth on the unspoiled end, i.e.\ reaction \eq{react3}.
The third term comes from reaction \eq{react2} and the fourth
term comes from reaction \eq{react4} and enters here and also in
\Eqs{lambdaL}{lambdaR} as a loss term.
For completeness, we note that the corresponding gain enters in
the evolution equations
\begin{equation}
{\dd[RL_{n}R]\over\dd t}=k_I[R_1][L_nR],\quad
{\dd[LR_{n}L]\over\dd t}=k_I[L_1][R_nL],
\label{RLnR+LRnL}
\end{equation}
which are not explicitly required for constructing a solution,
because these polymers no longer react with the monomers.
Nevertheless, solving \Eq{RLnR+LRnL} simultaneously with
\Eqs{Ln_new2}{Rn_new2} and \Eqss{L1R1}{RLevol} can be useful for monitoring
the evolution of the net chirality; see \Sec{Conservation}.

Note that, in contrast to the equations given by Sandars (2003), the
truncation levels for the terms $[L_n]$, $[L_nR]$, and $[RL_nR]$ are
here the same,
i.e.\ $n\leq N$ for both terms, whereas in the model of Sandars the longest
$L_nR_1$ chain has $n=N-1$, and the longest $R_1L_nR_1$ chain has only
$n=N-2$.
The reason we need to keep the same truncation levels for all three types
of polymers is that we want to ensure that the behavior near the end of
the chain ($n=N$) does not deviate from the behavior elsewhere ($n<N$);
see the discussion in \Sec{HomochiralGrowth}.
For example, to ensure continuous behavior of $[L_n]$ at $n=N$ we need
to keep the term $-2k_I[L_N][R_1]$ in \Eq{Ln_new2}.
This term, however, is the loss resulting from the gain of $[L_NR]$, so
we have to keep the evolution equation for this term as well.
Furthermore, the evolution equation for this term involves, in turn,
the term $-k_I[R_1][L_NR]$, which is the loss corresponding to the gain
of $[RL_NR]$.
If one regards the truncation level $N$ as an unrealistic feature of the
model, as we do, then all three polymer types should be truncated
at the same level.

\begin{figure}[t!]\begin{center}
\includegraphics[width=\textwidth]{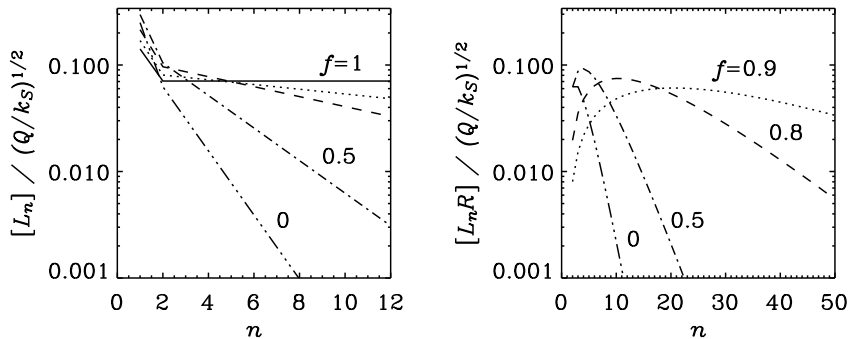}
\end{center}\caption[]{
$[L_n]$ (left) and $[L_nR]$ (right) of equilibrium solutions for different
values of $f$.
For $f=1$ we have $[L_nR]=0$, which cannot be seen in the logarithmic
representation.
}\label{pspectra}\end{figure}

In \Fig{pspectra} we show $[L_n]$ and $[L_nR]$ for a number of
equilibrium solutions for different values of $f$ and $k_I/k_S=1$.
The corresponding values of $[R_n]$ and $[R_nL]$ are small and
not shown, except when $f=0$ in which case the solution is fully
racemic with $[R_n]=[L_n]$ and $[R_nL]=[L_nR]$ and is simply
\begin{equation}
[L_nR]=(n-1)[L_n]=(n-1)2^{-n}[L_1].
\label{RacemicSolution}
\end{equation}
For $f=1$ the solution is given by \Eq{HomochiralEquilib}.

For $k_I/k_S=0.1$ the results are similar to those for $k_I/k_S=1$
provided $f>0.8$.
For $f<0.7$, however, the solution is fully racemic and therefore
the curves are independent of $f$.
This racemic solution is similar to the case
$k_I/k_S=1$ and $f=0.8$ that is shown in \Fig{pspectra}.

\section{Stability of the racemic equilibrium}
\label{SStability}

A realistic model of homochirality must also have an achiral (racemic)
equilibrium solution.
It is generally anticipated that this racemic solution can be destabilized
in the presence of catalytic reactions (Frank, 1953;
Avetisov and Goldanskii, 1993).
If the probabilities of adding left and right handed monomers to a
homochiral polymer are equal, i.e.\ if $k_I=k_S$, the racemic solution
given by \Eq{RacemicSolution} is also a possible solution for other
values of the fidelity than $f=0$, but it may of course be unstable.

\begin{figure}[t!]\begin{center}
\includegraphics[width=\textwidth]{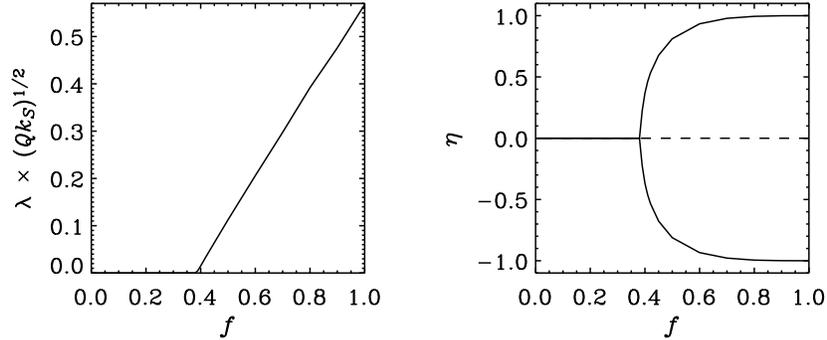}
\end{center}\caption[]{
Growth rate (left) and bifurcation diagram
showing a classical pitchfork bifurcation (right)
as a function of fidelity for $k_I/k_S=1$, and $N=50$.
The dashed line indicates an unstable solution.
}\label{bifurc}\end{figure}

We have carried out a numerical stability analysis by adding a small
($10^{-5}$) relative perturbation to the value of $[L_1]$ of the racemic
solution.
It turns out that for certain values of the fidelity $f$ the departure
of $[L_1]$ from the racemic equilibrium solution, $\delta[L_1]$, growth
exponentially in time like $e^{\lambda t}$.
In \Figs{bifurc}{bifurc01} we plot $\lambda$ obtained from the slope
of the graph of $\ln\delta[L_1](t)$ during the exponential growth phase,
i.e.\ before a new nonlinear equilibrium is attained.
In \Figs{bifurc}{bifurc01} we also plot the corresponding
chiral polarization parameter, $\eta$, as a function of $f$.
Here we have chosen to define $\eta$ as
\begin{equation}
\eta=\left(E_R-E_L\right)/\left(E_R+E_L\right).
\label{eta}
\end{equation}
It turns out that for $k_I/k_S=1$ the racemic solution is unstable when
$f>0.39$, and for $k_I/k_S=0.1$ it is unstable when $f>0.735$.
The transition from an achiral to a chiral solution is a typical
example of a pitchfork bifurcation; see \Figs{bifurc}{bifurc01}.
This result is in qualitative agreement with Sandars (2003) who
found that for $k_I/k_S=1$ the critical value of $f$ is around 0.21.
The differences in the numerical values are explained by differences
in the model (e.g., the coupling to the substrate and the length
of the maximum polymer length).

\begin{figure}[t!]\begin{center}
\includegraphics[width=\textwidth]{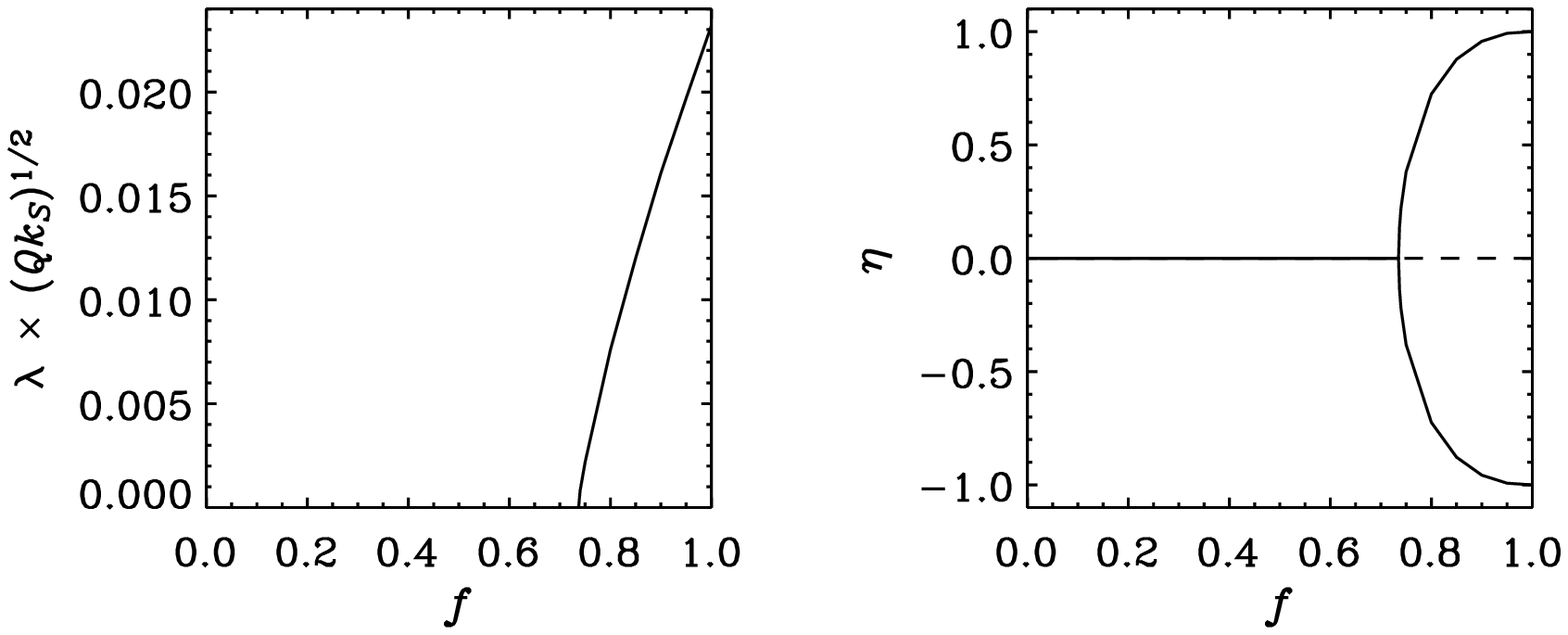}
\end{center}\caption[]{
Same as \Fig{bifurc}, but for $k_I/k_S=0.1$.
}\label{bifurc01}\end{figure}

The growth rate of the instability is important for determining the time
it takes for an almost racemic solution to become homochiral
(or at least non-racemic for $f\neq1$).
When $k_I/k_S=1$, the growth rate $\lambda$ is around 0.5, but it becomes
significantly smaller when the value of $k_I$ is reduced.
This shows explicitly that homochirality emerges as being due to
enantiomeric cross-inhibition.

\section{Conservation of chirality}
\label{Conservation}

For homochiral growth the relevant conservation law is given by
\Eq{dELdt} for $E_L$, and similarly for $E_R$.
In general, however, because of the interaction with left and
right handed monomers, there are no longer separate conservation
laws for $E_L$ and $E_R$.
Instead, the complete set of equations, \Eqs{Ln_new2}{Rn_new2}
together with \Eqss{L1R1}{lambdaR}, satisfies
\begin{equation}
{\dd\over\dd t}\Delta\tilde{E}=\Delta Q-\Delta\Lambda,
\label{H_cons}
\end{equation}
where $\Delta Q=Q_R-Q_L$ and $\Delta\Lambda=\Lambda_R-\Lambda_L$
are the net input and output rates of chirality, respectively, and
$\Delta\tilde{E}=\tilde{E}_R-\tilde{E}_L$ is the total chirality, where
\begin{equation}
\tilde{E}_R=\sum_{n=1}^N n[R_n]+\sum_{n=2}^N(n-1)[R_nL]
+\sum_{n=3}^N(n-2)[LR_nL],
\label{ERtilde}
\end{equation}
\begin{equation}
\tilde{E}_L=\sum_{n=1}^N n[L_n]+\sum_{n=2}^N(n-1)[L_nR]
+\sum_{n=3}^N(n-2)[RL_nR],
\label{ELtilde}
\end{equation}
denote the total numbers of right and left handed building blocks
(or enantiomers),
where opposite enantiomers are counted such that they annihilate
enantiomers of the opposite chirality.
The loss terms resulting from the finite truncation level, $N$, are
denoted by
\begin{equation}
\Lambda_R=2k_S N[R_1][R_N]+k_S(N-1)[R_1][R_NL],
\end{equation}
\begin{equation}
\Lambda_L=2k_S N[L_1][L_N]+k_S(N-1)[L_1][L_NR].
\end{equation}
In order to evaluate the quantities $\tilde{E}_R$ and
$\tilde{E}_L$ we have to integrate the evolution equations \eq{RLnR+LRnL}
for the production of terminally spoiled polymers -- even though they
undergo no further evolution.
In a sense the integration of the terminally spoiled polymers acts
only as counters that keeps track of the number
of polymers that are lost during the polymerization process.

The expressions for $\tilde{E}_R$ and $\tilde{E}_L$ involve sums over
$[LR_nL]$ and $[RL_nR]$, but since these quantities do not occur on the
right hand sides of the governing evolution equations, their values are
not constrained by the dynamics and depend on the initial conditions
and continue to evolve in time even though the system may have reached
an equilibrium.
The so defined net chirality can therefore not be used to characterize
a particular solution, and we have to restrict ourselves either to
$E_R$ and $E_L$, or to $\hat{E}_R$ and $\hat{E}_L$, which are defined
by taking only the first two sums in \Eqs{ERtilde}{ELtilde}, i.e.\
\begin{equation}
\hat{E}_R=\sum_{n=1}^N n[R_n]+\sum_{n=2}^N(n-1)[R_nL],
\label{ERtilde}
\end{equation}
\begin{equation}
\hat{E}_L=\sum_{n=1}^N n[L_n]+\sum_{n=2}^N(n-1)[L_nR].
\label{ELtilde}
\end{equation}
In \Tab{Tres} we list the resulting values of $\eta$,
defined in \Eq{eta}, an analogously defined
$\hat\eta=(\hat{E}_R-\hat{E}_L)/(\hat{E}_R+\hat{E}_L)$,
$\Delta E=E_R-E_L$, and $\Delta\hat{E}=\hat{E}_R-\hat{E}_L$.
We also give the mean polymer lengths, $N_R=\sum n[R_n]/\sum[R_n]$
and $N_L=\sum n[L_n]/\sum[L_n]$, of right and left handed polymers.

\begin{table}[t!]\caption{
Numerical results for $\eta$, $\hat\eta$, $\Delta E$, and $\Delta\hat{E}$
for different values of $f$.
The $\pm$ indicates that these values can have either sign.
The last column gives the typical value of $N$ necessary for obtaining
a converged result representing the limit $N\to\infty$.
For $f=1$ the results for $\Delta E$ ($=\Delta\hat{E}$) and $N_R$
do not converge
and we give the analytic expression for arbitrary $N$ instead.
}\vspace{12pt}{\begin{tabular}{cccccccc}
$f$ & $\pm\eta$ & $\pm\hat\eta$ & $\pm\Delta E$ & $\pm\Delta\hat{E}$
& $N_R$ & $N_L$ & $N$ \\
\hline
1 & $1$ & $1$ & \multicolumn{2}{c}{$\quarter[N(N+1)+2]/N^{1/2}$}
& $\half N+1/(N+1)$& -- & $N$\\
0.9 & $0.999$ & $0.9999$& $30.61$ & $143.73$ &19.0 & 1.0 & 500 \\
0.8 & $0.995$ & $0.9986$& $10.28$ & $45.06$  & 9.0 & 1.1 & 200 \\
0.7 & $0.978$ & $0.993$ & $5.170$ & $20.99$  & 5.6 & 1.1 & 100 \\
0.6 & $0.933$ & $0.975$ & $2.949$ & $11.01$  & 3.9 & 1.2 & 100 \\
0.5 & $0.813$ & $0.907$ & $1.648$ & $5.597$  & 2.8 & 1.2 &  50 \\
0.4 & $0.368$ & $0.482$ & $0.491$ & $1.500$  & 1.9 & 1.5 &  50 \\
0.38& $0$     & $0$     & $0$     & $0$      & 1.7 & 1.7 &  20 \\
\label{Tres}\end{tabular}}\end{table}

\section{Comparison with other models}

The polymerization model of Sandars (2003) is significantly different
from all the previously proposed models of homochirality that ignore
the detailed polymerization process by only describing
some scalar quantities, say $x$ and $y$, that are representative
of the number of left and right handed polymers.
In the papers by Saito and Hyuga (2004) it was shown that neither linear
nor nonlinear autocatalytic behavior suffice to produce homochirality,
and that a backreaction term is needed.
Their model equations are
\begin{eqnarray}
\begin{array}{l}
\dot{x}=x^2(1-r)-\epsilon x\cr
\dot{y}=y^2(1-r)-\epsilon y
\end{array}
\quad\mbox{(SH model)}
\label{SaitoHyuga}
\end{eqnarray}
where $r=x+y$ and $\epsilon$ is the feedback parameter.
For $\epsilon=0$ there is a continuous range of solutions along the line
$r\equiv x+y=1$, i.e.\ homochirality does not emerge unless the initial
condition is already homochiral.
For finite (but small) values of $\epsilon$ there are two nontrivial
stable fixed points.
(The trivial solution, $x=y=0$, is always a stable fixed point in
this model.)

The model of Saito and Hyuga (2004), hereafter the SH model,
does capture the expected behavior,
but it remains unsatisfactory in that its functional form has been introduced
ad hoc.
It is therefore desirable to derive simple model equations based
on the polymerization equations of Sandars (2003).
It turns out that, without changing the basic properties of the model,
a minimal version is still meaningful for $N=2$,
and that the equations for the semi-spoiled polymers, $[L_2R]$ and
$[R_2L]$, can be ignored (as already done by Sandars).
Thus, we only solve \Eqs{Ln_new2}{Rn_new2} together with \Eqss{L1R1}{lambdaR}.
Following Sandars (2003), we also assume that $C_L=[L_2]$ and $C_R=[R_2]$
(instead of $C_L=E_L$ and $C_R=E_R$, which would yield more complicated
expressions).
A further simplification can be made by regarding $[L_2]$ as a rapidly
adjusting variable that is enslaved to $[L_1]$ (and similarly for $[R_2]$).
This technique is also known as the adiabatic elimination of rapidly
adjusting variables (e.g., Haken, 1983).
\EEq{Ln_new2} becomes
\begin{equation}
0=k_S[L_1]^2-2[L_2]\Big(k_S[L_1]+k_I[R_1]\Big),
\end{equation}
which is solved for $[L_2]$ (and similarly for $[R_2]$), which in turn
couples back to the equations for $[L_1]$ and $[R_1]$ via $Q_L$ and $Q_R$.
Finally, we also treat the substrate $[S]$ as a rapidly adjusting variable,
i.e.\ we have $k_C[S]=Q/([L_2]+[R_2])$.
We emphasize that the adiabatic elimination does not affect the
accuracy of steady solutions.
It is convenient to introduce new dimensionless variables,
\begin{equation}
x=[R_1](2k_S/Q)^{1/2},\quad
y=[L_1](2k_S/Q)^{1/2},\quad
\tau=t(Qk_S/2)^{1/2}.
\end{equation}
In order to compare first with the SH model we restrict ourselves
to the special case $k_I/k_S=f=1$, which leads to the revised model
equations
\begin{eqnarray}
\begin{array}{l}
\dot{x}=x^2/\tilde{r}^2-rx,\cr
\dot{y}=y^2/\tilde{r}^2-ry,
\end{array}
\label{modeleqn}
\end{eqnarray}
where dots denote derivatives with respect to $\tau$ and
$r=x+y$ and $\tilde{r}^2=x^2+y^2$ has been introduced for brevity.
Equations \eq{modeleqn} resemble the equations of the SH model
in that both have a quadratic term proportional to $x^2$ (or $y^2$),
which is quenched either by a $1-r$ factor (in the SH model) or by
a $1/\tilde{r}^2$ factor in our model.
Furthermore, both models have a backreaction term proportional to $-x$
(or $-y$), but the coefficient in front of this term
($\epsilon$ in the SH model) is not constant but equal to $r$.

\begin{figure}[t!]\begin{center}
\includegraphics[width=\textwidth]{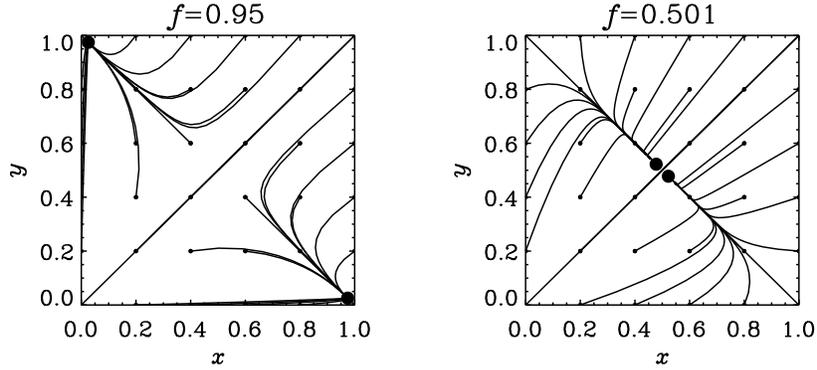}
\end{center}\caption[]{
Phase diagram showing the trajectories of solutions
of \Eq{modeleqnf} for two different values of $f$.
The starting points of each trajectory are marked by small dots
and stable fixed points are marked by big dots.
}\label{pruns051}\end{figure}

In the general
case with $k_I/k_S\neq1$, $f\neq1$, as well as finite values of
$C_{0x}=C_{0R}(2k_S/Q)^{1/2}$, and $C_{0y}=C_{0L}(2k_S/Q)^{1/2}$,
the equations read
\begin{eqnarray}
\begin{array}{l}
\dot{x}=(p\tilde{x}^2+q\tilde{y}^2+C_{0x})/\tilde{r}^2-r_x x,\cr
\dot{y}=(p\tilde{y}^2+q\tilde{x}^2+C_{0y})/\tilde{r}^2-r_y y,
\end{array}
\label{modeleqnf}
\end{eqnarray}
where we have introduced the abbreviations
$r_x=x+yk_I/k_S$, $r_y=y+xk_I/k_S$,
$\tilde{x}^2=x^2/2r_x$, $\tilde{y}^2=y^2/2r_y$,
$\tilde{r}^2=\tilde{x}^2+\tilde{y}^2+C_{0x}+C_{0y}$,
$p=(1+f)/2$, and $q=(1-f)/2$.

In \Fig{pruns051} we show trajectories of solutions of \Eq{modeleqnf}
for two different values of $f$ in an $(x,y)$ phase diagram.
Note that all equilibrium solutions lie on the line $r=1$.
This property allows us to calculate equilibrium solutions for
general values of $f$.
Inserting $y=1-x$ yields a cubic equation of which one solution
is always $x=1/2$.
This reduces the problem to a quadratic equation with the solution
\begin{eqnarray}
x=\half\left\{
\begin{array}{ll}
1\pm\sqrt{2f-1}\quad&\mbox{for $f\geq1/2$},\\
1 & \mbox{otherwise}.
\end{array}
\right.
\end{eqnarray}
Linearizing the equations around the racemic solution, $x=y=1/2$, yields
the growth rate
\begin{equation}
\lambda=2f-1.
\end{equation}
In agreement with our numerical results for large values of $N$,
this equation gives a linear dependence of the growth rate on
the fidelity.
This result also shows that for $f<1/2$ perturbations decay exponentially.

\begin{figure}[t!]\begin{center}
\includegraphics[width=.8\textwidth]{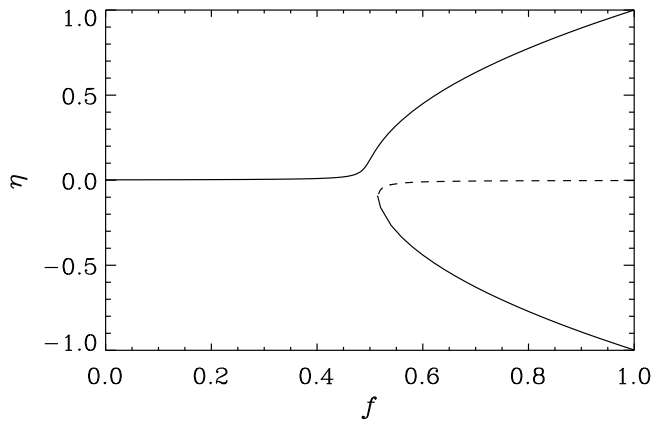}
\end{center}\caption[]{
Imperfect bifurcation obtained by solving \Eq{modeleqnf} for
$C_{0x}=0.001$ and $C_{0y}=0$ using the Newton-Raphson method.
}\label{pbranches}\end{figure}

In the presence of a biased, non-catalytic generation of monomers
(finite $C_{0x}$ or $C_{0y}$ with $C_{0x}\neq C_{0y}$) there is no
longer a perfectly racemic equilibrium solution.
The sign of $\eta$ for the solution for $f=0$ depends on the
sign of $C_{0x}-C_{0y}$.
Along this solution branch $\eta$ goes further away from zero in
a continuous fashion until $f=1$.
At some value of $f$ a pair of new solutions emerges, one is stable and
the other one unstable, but both have the opposite sign of $\eta$;
see \Fig{pbranches}.
Among these new branches, the stable one can only be reached via a
finite amplitude perturbation.
This behavior is called an imperfect bifurcation and has long been
anticipated in this context (Kondepudi and Nelson, 1983;
Kondepudi \ea 1986; Goldanskii and Kuzmin, 1989).

The steady solutions shown in \Fig{pbranches}
have been obtained by solving \Eq{modeleqnf} using the
Newton-Raphson method.
This method allows us to find both stable and unstable solutions.
Near the bifurcation point the diagram is extremely sensitive to the
addition of a bias parameter.
It is remarkable that for a value as small as $C_{0x}=10^{-3}$ a relatively
large gap has been produced in the bifurcation diagram.

Finite values of $C_{0x}$ and $C_{0y}$ could result from physical
influences, for example polarized synchrotron radiation from neutron stars
(but see Bonner, 1999), UV radiation in star-forming regions (Bailey, 2001),
or the parity violation of the electroweak force (e.g., Hegstrom, 1984). 
In all these cases the expected effect is however very small (Bada, 1995).
We emphasize, however, that the main reason for homochirality is the
instability of the racemic (or nearly racemic) solution, which is
hardly modified by a finiteness of $C_{0x}$ or $C_{0y}$.

\section{Conclusions}

The origin of homochirality has long been thought to be the result
of a bifurcation process that can vastly amplify a very small random
enantiomeric excess which can then prevail forever.
Generic model equations reproducing the expected bifurcation behavior
have so far mostly been proposed on an ad hoc basis.
It was therefore difficult to establish a connection between model
and reality.
According to the work of Saito and Hyuga (2004) one expects two
effects to be important: nonlinearity and backreaction.
However, the functional form of these terms remained open.
Furthermore, the meaning of non-perfect catalytic fidelity
and enantiomeric cross-inhibition within the framework of the
model were not clear.
In the present paper we have established a direct connection between
the more detailed polymerization model of Sandars (2003) and
the simpler model equation approach with only two ordinary differential
equations.
In particular, the present work has confirmed that the relevant
nonlinearity is indeed quadratic (as in the SH model), but it is not
quenched like $1-r$, but rather like $1/\tilde{r}^2$, where $r$ and
$\tilde{r}$ are measures of the total concentrations of monomers
(both right and left handed).
Furthermore, the feedback coefficient is not a small constant, as in
the SH model, but it is itself proportional to $r$.
More importantly, imperfect fidelity and enantiomeric cross-inhibition,
as well as the effects of a weakly biased non-catalytic production of
new monomers, have a quantitative meaning within the framework of
the reduced model.

For a more quantitative comparisons of the polymerization process with
experiments the full set of equations of Sandars (with the revisions
discussed above) is to be preferred.
A number of features that can only be captured by the full model.
An example is the wave-like propagation in the distribution of homochiral
polymers.
An experimental confirmation would help to quantify the growth coefficient
$k_S$ characterizing the probability that a polymer grows by a monomer
of the same chirality.
On the other hand, the growth coefficient for enantiomeric
cross-inhibition, $k_I$, determines primarily the minimum fidelity
parameter, $f$, above which bifurcation and hence homochiral growth is
at all possible.
It is indeed quite remarkable, that the main reason homochiral
growth occurs is that binding with a wrong enantiomer spoils further
polymerization on the corresponding end of the chain.
This leads to competition which is always a key feature of natural
selection processes such as these.

Homochirality in living organisms is a singular phenomenon.
Non-living chemical systems do in general not have a preferred
chirality.
In the models presented in this paper this is reflected in
\Figs{bifurc}{pbranches}.
The region of the phase diagram displaying homochirality is
characterized by high fidelity, i.e.\ high auto-catalytic accuracy.
The fidelity is expected to be significantly higher in living 
systems.
When an organism dies the auto-catalytic polymerization
stops and as a consequence the fidelity is sharply decreased.
The characteristic behavior of the polymerization changes from the chiral
to the racemic region of the phase diagram. The relaxation
of the system from the homochiral to the racemic state is
often very slow. It was in fact suggested by Hare and Mitterer 
(1967) and later by Bada \ea (1970)
that racemization of amino acids in fossil material could 
be used as a dating method.
Unfortunately it has turned out that the rate of racemization 
is strongly temperature dependent, which tends to make this
dating method unreliable.

\end{article}
\end{document}